\def\year{2021}\relax
\documentclass[letterpaper]{article} % DO NOT CHANGE THIS
\usepackage{aaai21}  % DO NOT CHANGE THIS
\usepackage{times}  % DO NOT CHANGE THIS
\usepackage{helvet} % DO NOT CHANGE THIS
\usepackage{courier}  % DO NOT CHANGE THIS
\usepackage[hyphens]{url}  % DO NOT CHANGE THIS
\usepackage{graphicx} % DO NOT CHANGE THIS
\usepackage{natbib}  % DO NOT CHANGE THIS AND DO NOT ADD ANY OPTIONS TO IT
\usepackage{caption} % DO NOT CHANGE THIS AND DO NOT ADD ANY OPTIONS TO IT
\title{SenTag: a Web-based Tool for Semantic Annotation of Textual Documents}
\title{SenTag: a Web-based Tool for Semantic Annotation of Textual Documents}
\author {
    Andrea Loreggia,\textsuperscript{\rm 1}\thanks{ A.\ Loreggia  has been supported by the H2020 ERC Project ``CompuLaw'' (G.A.\ 833647).}
    Simone Mosco, \textsuperscript{\rm 2}
    Alberto Zerbinati \textsuperscript{\rm 2} \\
}
\affiliations {
    \textsuperscript{\rm 1} European University Institute \\
    \textsuperscript{\rm 2} University of Padova \\
    andrea.loreggia@gmail.com, simone.mosco@studenti.unipd.it, alberto.zerbinati@studenti.unipd.it
}

\begin{document}

\maketitle

\begin{abstract}
In this work, we present SenTag, a lightweight web-based tool focused on semantic annotation of textual documents. The platform allows multiple users to work on a corpus of documents. 
The tool enables to tag a corpus of documents through an intuitive and easy-to-use user interface that adopts the Extensible Markup Language (XML) as output format. The main goal of the application is two-fold: facilitating the tagging process and reducing or avoiding for errors in the output documents. 
Moreover, it allows to identify arguments and other entities that are used to build an arguments graph. It is also possible to assess the level of agreement of annotators working on a corpus of text.
\end{abstract}

\section{Introduction}
In the scope of Natural Language Processing (NLP), many applications necessitate of a great amount of annotated documents in order to train machine learning (ML) systems. Group of experts are employed to manually annotate documents in order to identify important details that can be used to train artificial intelligence models \cite{poudyal2020echr}. In this preliminary step, a predefined structures language is adopted \cite{lippi2019claudette}, such as Extensible Markup Language (XML) or JSON.
%As in many other ML approaches, input data cannot be used as it is but it has to be managed in order to identify important information included in the text \cite{poudyal2020echr}. 
%For instance, tagging the name of the judge or the part of the sentence which identify the request of the counter-part. 
Manually annotated datasets play an important role in the definition of gold standards. For instance, such datasets might be used for training machine learning tools to extract argument-related information from case texts \cite{ashley2017artificial}, to predict the outcome of a sentence (e.g., \cite{DBLP:conf/jurix/MedvedevaXWV20}), or to train machine learning systems able to summarize the original text (e.g., \cite{DBLP:conf/jurix/XuSA20}). 
Unfortunately, the manual process is prone to errors. For instance, annotators may introduce non-existing or misspelled tags. Thus, annotators have to spend time for checking the consistency and the validity of their work. 
In this work, we present SenTag, a web-based application %based on the paradigm of "What You See Is What You Get" (WYSIWYG), 
that provides an intuitive interface for document annotation.
The application allows multiple users to work on the same corpus of documents. Users can belong to 3 different groups (i.e., admins, editors, and annotators) based on the role that each user should play. Users that belong to the admin group are in charge of creating other users called editors and annotators. Editors upload XML schemas, documents, and specify which schema an annotator will use to annotate a document. At the end of the tagging phase, an annotator has to validate an annotated document against the XML schema. In case of errors (e.g., attributes with missing values), the annotator will be alerted. Otherwise, the resulting document complies with the schema and thus the tagging process produced a well-formed document.
Moreover, the application allows the use of specific tags to identify and easily build a graph of arguments: annotators can use these tags to identify different arguments in a document and define relationships among them. To the best of our knowledge, no existing platform implements all these features altogether.

\textbf{Related Work.}
The use of machine learning techniques shows their potential in getting better performance in this area \cite{slonim2021autonomous}, for instance in tasks such as document analytic or argument mining. These systems still strongly depend on manually tagged datasets and on the ability of human annotators. Unfortunately, obtaining good manual annotations is not an easy task. %Most of the time, documents are assigned to group of experts who annotate independently the corpus of texts. Their outputs might be compared to evaluate the level of agreement. 
To facilitate this task, some systems were developed over the last few years. Due to lack of space, we report a non-exhaustive list of works and we point the reader to a complete review on the topic, such as \cite{ashley2017artificial}): Gloss is an annotation system developed by researchers from the University of Pittsburgh, it leverages on some individual components to assist the user during the whole annotation process, including corpus assembly, type system definition, document annotation, as well as quality control. NER Annotator\footnote{https://github.com/tecoholic/ner-annotator - Last accessed 27th July 2021} is another web-based tool which provide a graphical user interface that helps users to annotate documents and generates traning data as a JSON which can be readily used. Unstructured Information Management applications (UIMA) is a suite of software systems developed to analyze large volumes of unstructured information \cite{UIMA:FERRUCCI:2003}. The suite provides a developer’s toolbox software that also includes an annotation interface. 
GATE Teamware is an open-source, web-based, collaborative text annotation framework. It enables users to carry out complex corpus annotation projects, involving distributed annotator teams \cite{bontcheva2013gate}. WebAnno is a generic web-based annotation tool for
distributed teams \cite{de2016web}.

\section{SenTag}
SenTag is developed in Python 3.9 as a Django 3.2 app employing VUE 3 for the graphical interface. 
The application is based on NER Annotator, from which it derives and expands the tagging part. 
%It is available for demo or test at the following link \url{DEMO URL}. The application can be accessed using the user XXX.
The main strengths of the application are: a) security and user management, b) multiple annotators and agreement score, c) intuitive interface for text annotation, d) graph of arguments.
All these features allow to build XML documents avoiding annotators for a deep understanding of the language and thus reducing possible errors.

\textbf{Security and User Management.}
The application allows to group users into three different categories:
%\begin{itemize}
a) \textbf{admin}: users with the highest level of rights. Admins can create other users and assign them to a specific group. They also have the rights to do all the tasks allowed to editors;
b) \textbf{editor}: this type of users is in charge of: (i) uploading texts and XML schema; (ii) assigning texts to the associated schema; (iii) assigning annotators to texts they should annotate; (iv) check for annotators agreement. Editors also have the rights to perform all the tasks allowed to annotators;
c) \textbf{annotator}: this type of users is in charge of tagging documents that are assigned to them using the set of tags specified in the schema associated to each document. Annotators can also validate their work against the XML schema after completing the annotation phase and build the graph of arguments based on the  arguments identified in the document.
%\end{itemize}

\begin{figure}
  \centering
  \includegraphics[width=\linewidth]{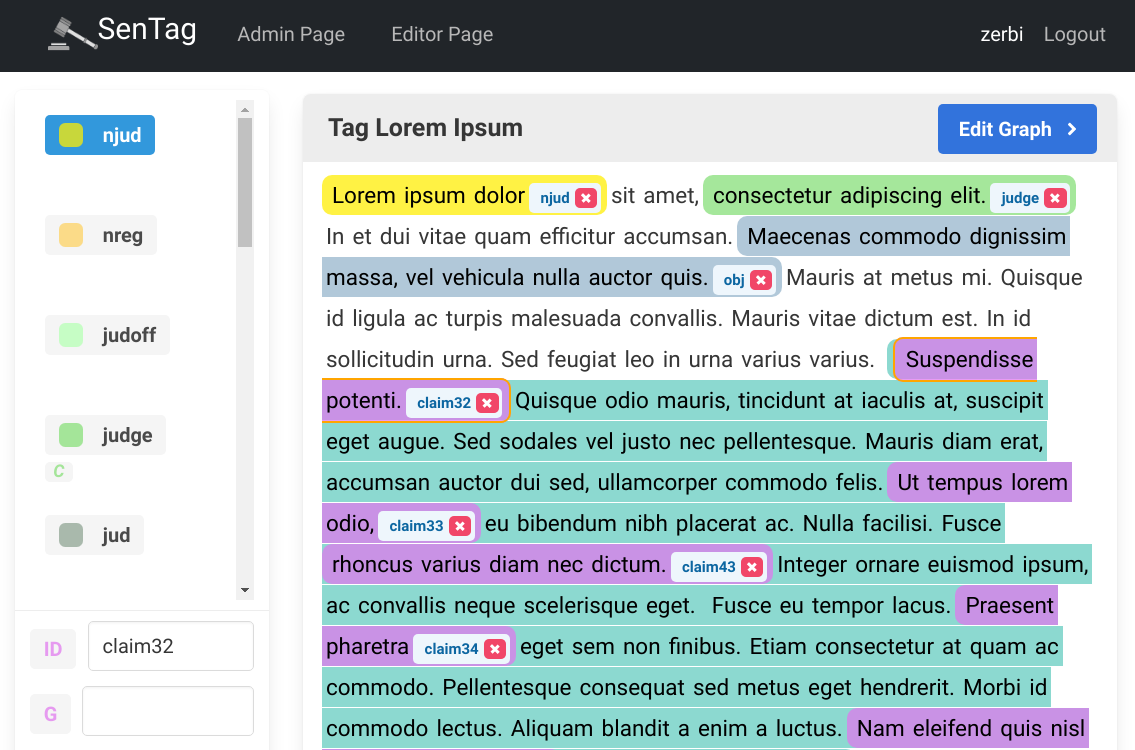}
  \caption{The annotation interface}
  \label{fig:annotation interface}
\end{figure}

%\subsubsection{Editors}
%Editors are in charge of uploading to the platform XML schema and documents that should be annotated. %Figure \ref{fig:editor page} reports a screenshot of an editor interface example. 
%Through this interface editors can upload XML schema and documents. They can also associate each document to the schema that has to be used for annotations. Moreover, they can also specify who (among all the annotators) is in charge for the annotation of each document in the corpus, check the agreement scores and download the annotated documents. 

%\begin{figure}
%  \centering
%  \includegraphics[scale=0.2]{img/screenshot editor page.png}
%  \caption{The editor interface}
%  \label{fig:editor page}
%\end{figure}

\textbf{Multiple Annotators and Agreement Score.}
Multiple annotators can be assigned to a document. In this case, the interface will report statics about the quality and the validity of the annotation performed on the document. For each document, the Krippendorff's alpha \cite{krippendorff2011agreement} is reported describing the level of agreement for the group of annotators working on the document. Moreover, for each annotator, the interface reports whether he/she has completed the annotation for a document. It also reports whether each annotated document passed the validity check against the XML schema.

\begin{figure}
  \centering
  \includegraphics[width=0.8\linewidth]{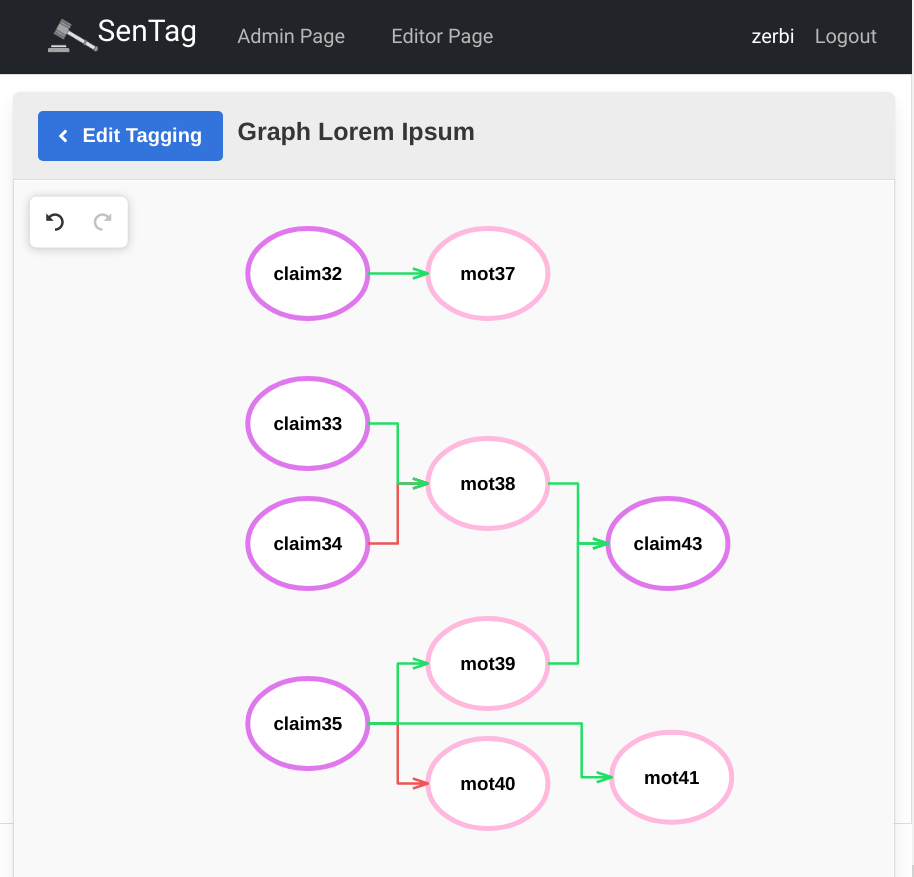}
  \caption{The graph interface}
  \label{fig:graph interface}
\end{figure}

\textbf{Text Annotation.}
Annotators are in charge of the annotation of documents. After being authenticated, an annotator is proposed with a list of documents assigned to him/her. Documents are grouped based on whether the annotation phase is completed and thus validated. The annotation is done using the interface depicted in Figure \ref{fig:annotation interface}. The screen is divided into two different areas. On the left side, all the available tags (and their attributes) are reported, this enables the annotator to choose which tag to use. When a tag is selected, the list of its attributes appears. This allows the annotator to change (if she wants) the value for a specific attribute or simply to consult it for future purposes. After clicking on a tag, the annotator can select part of the text which will be highlighted with the color associated to the tag. This would serve as a reminder for the annotator, the name of the tag associated to a part of the text is always visualized.

\textbf{Graph of Arguments.} The platform allows to build the graph of arguments from the tagged document. The XML schema might contains tags with the specific attribute GRAPH. When part of the text is tagged with one of these tags, automatically a node appears in the graph. A dedicated area allows annotators to draw edges between nodes in order to describe relationships among arguments. Any time a node is connected to another one, the list of ancestors and descendants of the nodes are adjusted with the ids of the selected nodes. This allows to enrich the final XML document by adjusting the correspondent attributes in the output file. Figure \ref{fig:graph interface} depicts the graphical interface for graph editing.

\section{Conclusion}
We presented SenTag a new lightweight web-based tool focused on semantic annotation of textual documents. The main goal of this tool is two-fold: on one side, the tool makes available an intuitive environment for the annotation of a corpus of text with a cooperative multi-user approach; on the other side, it avoids for a direct editing of the XML tags, minimizing the introduction of errors and mispelled output.

\bibliography{biblio} 

\end{document}